\documentclass[%
 reprint, superscriptaddress,%groupedaddress,%unsortedaddress,%runinaddress,%frontmatterverbose, %preprint,%preprintnumbers,%nofootinbib,%nobibnotes,%bibnotes, %pra,%prb,%rmp,%prstab,%prstper,
 showkeys,floatfix,amsmath,amssymb, aps,onecolumn,11pt
]{revtex4-2}
\usepackage{hyperref}
\hypersetup{
    colorlinks=true,
    linkcolor=blue,
    citecolor=red,
    filecolor=magenta,      
    urlcolor=cyan}

\usepackage{graphicx}% Include figure files
\usepackage{dcolumn}% Align table columns on decimal point
\usepackage{bm}% bold math
\usepackage{braket} %braket
\usepackage{xr}% to cross reference between document and SI
\usepackage{xcolor}
\usepackage{siunitx}
\makeatletter
\newcommand*{\addFileDependency}[1]{
  \typeout{(#1)}
  \@addtofilelist{#1}
  \IfFileExists{#1}{}{\typeout{No file #1.}}
}
\makeatother

\newcommand*{\myexternaldocument}[1]{
    \externaldocument{#1}
    \addFileDependency{#1.tex}
    \addFileDependency{#1.aux}
}

\myexternaldocument{SI}

\DeclareUnicodeCharacter{2212}{-}

\begin{document}
\preprint{APS/123-QED}

\title{
Mode-specific Coupling of Nanoparticle-on-Mirror Cavities with Cylindrical Vector Beams
}
\author{Valeria Vento}
\affiliation{%
  École polytechnique fédérale de Lausanne (EPFL), Institute of Physics, CH-1015 Lausanne, Switzerland 
}%
\author{Philippe Roelli}%
\affiliation{%
  CIC nanoGUNE, Nano-optics Group, E-20018 Donostia-San Sebastián, Spain
}%
\author{Sachin Verlekar}%
\affiliation{%
  École polytechnique fédérale de Lausanne (EPFL), Institute of Physics, CH-1015 Lausanne, Switzerland 
}%
\author{Christophe Galland}
\email{chris.galland@epfl.ch}
\affiliation{%
  École polytechnique fédérale de Lausanne (EPFL), Institute of Physics, CH-1015 Lausanne, Switzerland 
}%
\date{\today}% It is always \today, today,
             
%Use showkeys class option if keyword

\begin{abstract}

\textbf{Abstract }
Nanocavities formed by ultrathin metallic gaps, such as the nanoparticle-on-mirror geometry, permit the reproducible engineering and enhancement of light-matter interaction thanks to mode volumes reaching the smallest values allowed by quantum mechanics. Although a large body of experimental data has confirmed theoretical predictions regarding the dramatically enhanced vacuum field in metallic nanogaps, much fewer studies have examined the far-field to near-field input coupling. Estimates of this quantity usually rely on numerical simulations under a plane wave background field, whereas most experiments employ a strongly focused laser beam. Moreover, it is often assumed that tuning the laser frequency to that of a particular cavity mode is a sufficient condition to resonantly excite its near-field.
Here, we experimentally demonstrate selective excitation of nanocavity modes controlled by the polarization and frequency of the laser beam. We reveal mode-selectivity by recording fine confocal maps of Raman scattering intensity excited by cylindrical vector beams, which are compared to the known excitation near-field patterns. Our measurements allow unambiguous identification of the transverse vs. longitudinal character of the excited cavity mode, and of their relative input coupling rates as a function of laser wavelength.
The method introduced here is easily applicable to other experimental scenarios and our results are an important step to connect far-field with near-field parameters in quantitative models of nanocavity-enhanced phenomena such as molecular cavity optomechanics, polaritonics and surface-enhanced spectroscopies. 

\end{abstract}

\maketitle

%TO BE CITED:
% \cite{fulmes2018} Mapping the electric field distribution of tightly focused cylindrical vector beams with gold nanorings

%TO BE ADDED:
% (1) sentence on historical idea of Novotny + citation
% (2) sentence on usage of radially polarized light for Raman + citations
% (3) sentence on PL data in SI
% (4) note on local contrast of coupling achieved by switching electronically polarization (see thesis)

\pagebreak
\section{Introduction} Plasmonic nanocavities are capable of capturing and confining light in dimensions much smaller than the free-space wavelength, where optically active materials can be positioned. Light-matter interaction is thereby greatly enhanced; when its rate overcomes all intrinsic dissipation rates the strong-coupling regime can be reached \cite{chikkaraddy2016,xiong_room-temperature_2021, heintz_few-molecule_2021, liu_nonlinear_2021, bylinkin_real-space_2021}.
More broadly, plasmonic nanocavities have become instrumental in a number cutting-edge technologies: Tip-Enhanced Raman Spectroscopy (TERS) \cite{zhang2013a,lee_visualizing_2019}; chemistry \cite{sun_-situ_2012} and electroluminescence \cite{kern2015} at the single molecule limit; enhanced nonlinearities \cite{celebrano_mode_2015}; single-photon sources \cite{hoang_ultrafast_2016}; coherent frequency upconversion \cite{chen_continuous-wave_2021}.
A common challenge for most applications is to achieve efficient coupling between a travelling electromagnetic wave (the far-field) and a confined plasmonic nanocavity mode (the near-field) \cite{li2018, baumberg_extreme_2019}. To this aim, a precise knowledge of the spectral and spatial distributions of the plasmonic modes are required, as well as techniques to identify, tune and optimize the coupling from far field radiation to specific subwavelength cavity modes. 

Among many different geometries studied in the literature, nanoparticle-on-mirror (NPoM) structures \cite{aravind1982,lee2022} constitute versatile, robust, reproducible and easy-to-fabricate plasmonic nanocavities with intrinsic antenna functionality \cite{li2018, baumberg_extreme_2019}, enabling degrees of confinement and enhancement of optical fields reaching their fundamental limits \cite{chen2018}. The NPoM optical resonances have been classified as transverse waveguide (S) and antenna (L) modes, which typically mix to form hybridized (J) gap modes presenting distinct near- and far-field radiation patterns \cite{lassiter_plasmonic_2013, tserkezis2015a, chikkaraddy2017, kongsuwan_plasmonic_2020, elliott_fingerprinting_2022}. 
Rigorous quantization of these highly dissipative modes and definition of their mode volumes have also been developed in the framework of non hermitian quasi-normal modes \cite{hughes2021,wu_nanoscale_2021}. 

Near field intensities for all gap modes being tightly confined within the spacer material, direct and quantitative analysis via near-field scanning probe or electron energy loss spectroscopy has not been achieved. Consequently, experiments rely on elastic dark-field scattering spectra obtained under quasi-plane wave excitation to infer the nanocavity spectrum and which mode is excited at a particular laser wavelength \cite{li_observation_2020,chikkaraddy2017}. Not only is this approach underestimating or neglecting the contribution of `dark' modes (which do not couple well to incident plane waves but can be efficiently excited by scatterers or emitters in the near field), it is also unable to make prediction as to the relative input coupling rate of a strongly focused laser beam to a particular mode, which can depend not only on laser wavelength, but also on the near field polarization at the focus \cite{long_reproducible_2016, fulmes_mapping_2018, shang_characterizing_2019, grosche_towards_2020, tang_plasmonic_2021}.
Knowledge of the cavity input coupling rate is required to infer the intracavity excitation number, which in turns govern all optically-driven processes in various applications including SERS, photochemistry and photocatalysis, plasmon-enhanced luminescence, nonlinear optics and frequency conversion. Its role and physical meaning has been evidenced using various formalisms: indirectly via quantum master equations \cite{esteban_strong_2014,sanchez-barquilla_cumulant_2020}, in plasmon induced transparency \cite{liu_plasmonic_2009,adato_engineered_2013,neuman_importance_2015,neubrech_surface-enhanced_2017} and also in the context of molecular cavity optomechanics \cite{roelli2016,schmidt2016b,schmidt2017,roelli_molecular_2020,esteban_molecular_2022}, which aims at a more complete description of vibrational and plasmonic correlations and dynamics in SERS. Recent theoretical developments have moreover pointed out its importance for a correct estimate of dissipative coupling rates \cite{primo_quasinormal-mode_2020}.

%Experimentally, these resonances have been mapped using near-field scanning optical microscopy (NSOM) \cite{yang_mapping_2019, yanai_2014_near}, fluorescence scanning \cite{novotny_longitudinal_2001} or 
%These techniques are based on the detection of the light elastically scattered from the nanostructures or from fluorescence of molecules embedded in the gap, and therefore they only probe the far-field radiation, which is a partial signature of the plasmonic mode and can be insufficient for discrimination.

Here, we present a simple and efficient method to determine which nanocavity modes are effectively excited depending on adjustable laser wavelength and polarization. %and to characterise thoroughly the relative coupling efficiency of gap modes to incident waves, as a function of laser wavelength and polarization.
We record the SERS intensity from molecules embedded in NPoM structures under tightly-focused radially, azimuthally and linearly polarized excitation as a function of the NPoM position within the laser focus. 
%We show that such measurements constitute a simple Raman-based tool for plasmonic mode discrimination that is  sensitive to the near-field characteristics of the tightly confined modes in the nanogap. 
We show that the intensity of the SERS signal is typically dominated by the coupling between a single vectorial component of the incoming light field and a single plasmonic gap mode. 
%To this aim, we compare the measured Stokes signal of a certain vibrational mode to the simulated components of the selected vectorial beam. 
We demonstrate that polarization and wavelength tuning of the excitation beam allow to address specific ultraconfined plasmonic modes and optimize simultaneously their coupling to traveling light modes and their parametrically enhanced interaction with the gap material. Finally, we prove the flexibility of our technique on different plasmonic structures, whose resonances cannot be easily identified with elastic dark-field scattering.
{We anticipate that our method will help understanding, modelling and optimising mode-specific excitation and in-coupling efficiency for a broad range of nanocavities, thus boosting further the desired light-matter interaction while reducing unwanted effects such as Joule heating, and allowing to unlock the full potential of multimode photonic nanostructures.} 
%Alternatively, if the NPoM geometry and the supported modes are precisely known, our technique can be used as simple Raman-based tool for field reconstruction of tightly-focused vectorial beams, which has been carried out in the past with more complex methods based on the interference of elastically scattered light \cite{bauer_nanointerferometric_2014}.

%%%%%%%%%%%%%%%%%%%%%%%%%%%%%%%%%%%%%%%%%%%%%%%%%%%%%%%%%%%%%%%%%%%%%%%%%%%%%%%%%%%%%%%%%%%%%%%%%%%%%%%%%%%%%%%%%%%%%%%%%%%%%%
\begin{figure}[h!]
\centering
\includegraphics[width=0.7\textwidth]{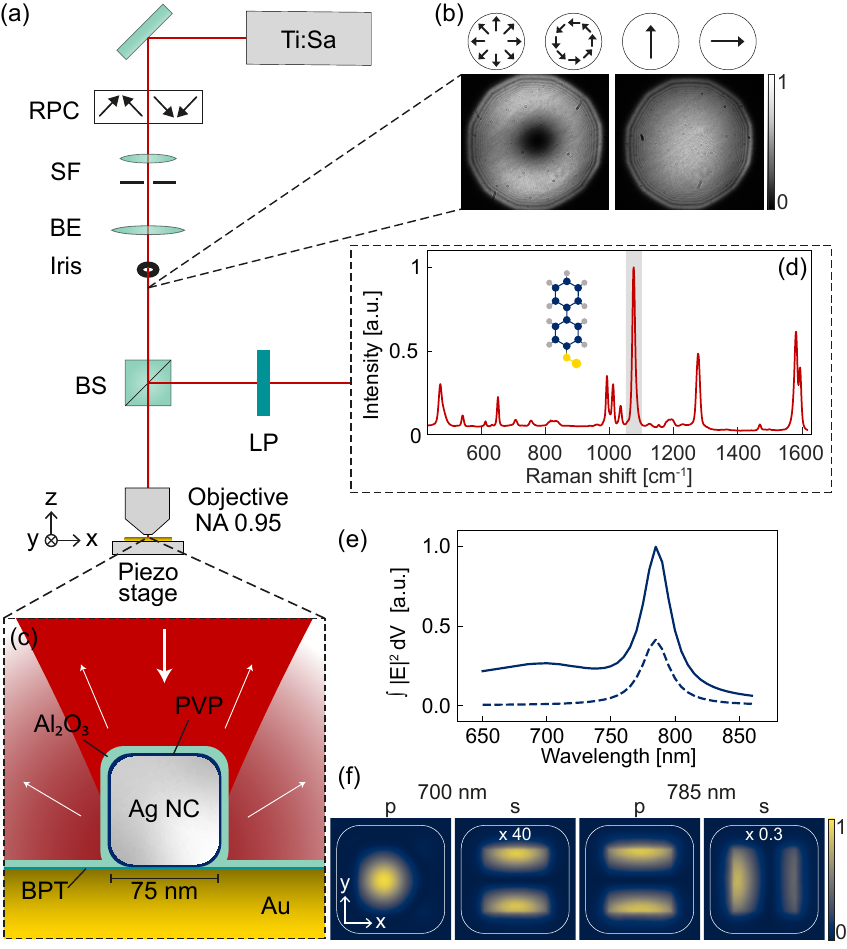}
  \caption{(a) Sketch of the experimental setup. RCP: Radial polarization converter; SP: spatial filter, BE: beam expander; BS: beam splitter; LP, long-pass filter. (b) Beam profiles after the iris for different laser polarizations.  (c) Sketch of a single nanocavity being scanned across the excitation focal plane. (d) Acquired Raman spectrum for a particular sample position; the Stokes scattering intensity from the vibrational mode at $\sim 1075$ cm$^{-1}$ is integrated over the gray region for subsequent construction of excitation maps. (e) FEM simulations of $|E|^2$ integrated over the gap region for $p$-polarized (solid line) and $s$-polarized (dashed line) plane-wave excitation (incidence angle $\theta\simeq\frac{\pi}{2}$). (f) Near-field intensity distribution along the $xy$ plane crossing the center of the gap for $700$-nm and $785$-nm excitation with $p$ and $s$ polarization. The simulations are realized for a gap size of $2.5$~nm and rounding factor of $0.44$ (see SI). The cube outline is indicated in white.}  
  \label{fig:setup}
\end{figure}
%%%%%%%%%%%%%%%%%%%%%%%%%%%%%%%%%%%%%%%%%%%%%%%%%%%%%%%%%%%%%%%%%%%%%%%%%%%%%%%%%%%%%%%%%%%%%%%%%%%%%%%%%%%%%%%%%%%%%%%%%%%%%%

\section{Results}
\paragraph*{Experimental setup. ---} 
A simplified description of the experimental setup is provided in Figure~\ref{fig:setup}a. On the laser path, a radial polarization converter followed by a spatial filter with a $20$-$\SI{}{\micro\meter}$ pinhole generates radially and azimuthally polarized light beams, which display doughnut-shape intensity profiles (Figure~\ref{fig:setup}b). The linearly-polarized Gaussian beam is obtained by bypassing the polarization converter. After a beam expander, an iris fixes the beam diameter to $5$~mm, which results in a filling factor $D \sim 0.7$ of the objective back aperture. The incoming field couples to the sample via a $0.95$-NA objective. The Stokes side of the inelastically scattered field is spectrally filtered before reaching the spectrometer.

The wavelength tunability of the Ti:Sapph laser source and the polarization modularity of our setup (see SI for details) enable a variety of different Raman excitation settings. The radial and azimuthal far fields correspond to superpositions of first-order Hermite-Gaussian modes ($\text{HG}_\text{mn}(x,y)$) $\boldsymbol{E}_\text{R}=\text{HG}_{10}(x,y)\boldsymbol{n}_x+\text{HG}_{01}(x,y)\boldsymbol{n}_y$ and $\boldsymbol{E}_\text{A}=\text{HG}_{10}(x,y)\boldsymbol{n}_y - \text{HG}_{01}(x,y)\boldsymbol{n}_x$, respectively, where $\boldsymbol{n}_i$ ($i=x,y$) are unit vectors perpendicular to the propagation direction. These travelling modes, solutions of the Helmholtz equation, belong to the set of cylindrical vector beams and are currently used in a variety of nano-optics applications \cite{zuchner_light_2011, rosales-guzman_review_2018}. Such beams feature position-dependent polarization and a singular point of zero intensity on the optical axis, giving them a characteristic doughnut-shaped intensity profile. In the same Hermite-Gaussian base, the linearly polarized Gaussian beams are defined as: $\boldsymbol{E}_\text{Lx}=\text{HG}_{00}(x,y)\boldsymbol{n}_x$ and $\boldsymbol{E}_\text{Ly}=\text{HG}_{00}(x,y)\boldsymbol{n}_y$.  

%the $J_-$ and $S_{11}$ modes appear for $690$-nm $p$-polarized excitation and $790$-nm $s$-polarized excitation respectively. 

%(b) SEM image of a nanocube with rounded edges. (c) FEM simulations of $|E|^2$ for a $p$-polarized (incidence angle $\theta\simeq0$) and $s$-polarized ($\theta=\pi/2$) plane-wave excitation.
 
% So far, Radially polarized beams have been considered for their ability to improve the SERS and TERS (tip-enhanced Raman scattering) sensitivity. A beam with radial polarization can be focused to a smaller spot size than a linearly polarized one; and more importantly, a stronger longitudinal field component is generated for the same incoming power \cite{quabis_focusing_2000, dorn_sharper_2003}, which is beneficial for the excitation of plasmonic modes polarized perpendicular to a substrate \cite{long_reproducible_2016,zhang_nanoscale_2010}.Here we take a more systematic and generic approach and show that the choice of the excitation settings strictly depends on the target mode, and therefore the employment of radial polarization in some spectral regions deteriorate the SERS sensitivity.

The sample is fixed on a piezo-stage that is finely scanned across the focal plane in $x$ and $y$ directions, for a fixed $z$ position along the optical axis (see SI for details). 
The Raman-active nanocavities are fabricated on a template-stripped $150$~nm thick gold film, on which biphenyl-4-thiol (BPT) molecules form a self-assembled monolayer (SAM) and $75$-nm silver nanocubes (NC) are dropcasted (Figure~\ref{fig:setup}c). The Polyvinylpyrrolidone (PVP) ligand attached to the cubes (NanoXact by nanoComposix) contributes to the gap size. %, which is estimated as $\sim 2$~nm$. 
A $5$~nm $\text{Al}_2\text{O}_3$ coating formed by atomic layer deposition (ALD) improves the stability of the nanostructures against oxidation and laser induced damage \cite{chen2018a}. 

At each sample position, a cavity-enhanced Raman spectrum of the BPT molecules is acquired, as illustrated in Figure~\ref{fig:setup}c. The three most intense peaks at $1075$ cm$^{-1}$, $1280$ cm$^{-1}$ and $1585$ cm$^{-1}$ correspond to the ring stretching, S-H stretching and C-H stretching vibrational modes, respectively. 
In the following, we use the integrated intensity over the ring stretching peak as a function of the $x,y$ coordinate of the sample scan to build near-field excitation maps (intensities of other vibrational modes are analysed in the SI). 
We used a large enough slit opening at the spectrometer input to ensure that the detection efficiency is independent of sample position over the range of the maps presented below. Consequently, we consider that the extracted intensity of the Raman signal is a faithful measure of the relative excitation efficiency when scanning the sample. 
We calibrated the wavelength- and polarization-dependent detection efficiency of the setup from the sample to the detector, as well as the  wavelength- and polarization-dependent transmission of the excitation path, see Figure~\ref{fig:AziHV} and accompanying discussion. 

%%%%%%%%%%%%%%%%%%%%%%%%%%%%%%%%%%%%%%%%%%%%%%%%%%%%%%%%%%%%%%%%%%%%%%%%%%%%%%%%%%%%%%%%%%%%%%%%%%%%%%%%%%%%%%%%%%%%%%%%%%%%%%
\begin{figure}[h!]
\centering
\includegraphics[width=0.7\textwidth]{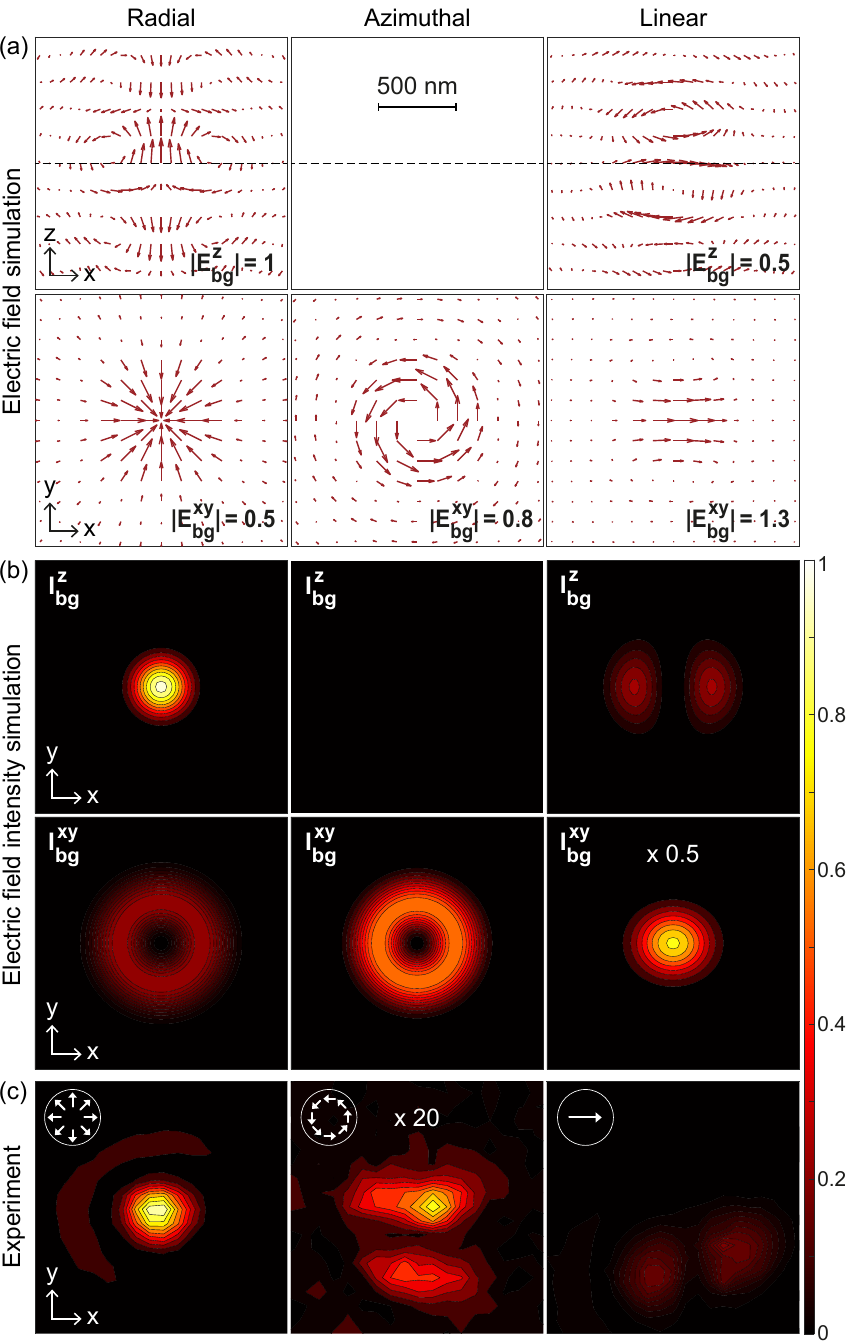}
  \caption{Comparison of experimental Raman intensity with calculated excitation fields near focus. (a) Computed background electric field in the $xz$ and $xy$ planes, where the dashed line represents the focal plane. In each panel, the maximum value of the longitudinal $E_\text{bg}^z(\boldsymbol{r},\epsilon)$ or transverse $E_\text{bg}^{xy}(\boldsymbol{r},\epsilon)$ field component is reported in relative units. (b) Corresponding longitudinal ($z$) and transverse ($r$) components of the focused field intensity $I_\text{bg}(\boldsymbol{r}_0,\epsilon)$ (across the focal plane $z=0$). All the simulations are realized by taking into account the specification of our objective, the filling factor $D$ and the experimental wavelength.
  (c) Measured maps of the Stokes intensity at $\SI{1075}{\centi\meter}^{-1}$ obtained with an excitation wavelength of $710$~nm (SEM image of the nanocube in Figure~\ref{fig:SEM-simul}). %\ref{fig:SEM-simul}. 
  Some panels are rescaled in intensity when indicated. } 
  \label{fig:maps-simul}
\end{figure}
%%%%%%%%%%%%%%%%%%%%%%%%%%%%%%%%%%%%%%%%%%%%%%%%%%%%%%%%%%%%%%%%%%%%%%%%%%%%%%%%%%%%%%%%%%%%%%%%%%%%%%%%%%%%%%%%%%%%%%%%%%%%%%

\paragraph*{Theory of polarization-dependent Raman signals. ---}
We briefly discuss how the enhancement of the Raman signal depends on the geometry of the nanocavity and on the excitation conditions (cf. near-field intensity simulations in Figure~\ref{fig:setup}e-f). The considered NPoM geometry is expected to present one hybrid mode $J_-$ at $\sim700$~nm and one transverse $S_{11}$ mode at $\sim785$~nm. The mode $J_-$ comes from the mixing of the transverse $S_{02}$ and longitudinal $L_{01}$ modes and has a prevalent longitudinal polarization \cite{zhang_surface_2019}. Note that the near-field polarization \textit{inside the nanogap} is mostly longitudinal (along $z$) for all modes. 
%Therefore, the coupling to the $J_-$ mode is maximal for $p$-polarized excitation at $700$~nm, while the intensity drops for $s$-polarized excitation, which couples weakly to the far detuned $S_{11}$. This polarization dependent coupling is consistent with previous dark field scattering studies \cite{chikkaraddy2017}. %We will show that the the coupling rate of the incoming field plays a fundamental role in the estimation of the Raman enhancement. 
As suggested by previous dark-field scattering studies \cite{chikkaraddy2017}, the coupling of this NPoM mode to a linearly polarized beam impinging at grazing angle would be optimal for $p$-polarization. Oppositely, $s$-polarized light would preferentially couple to the $S_{11}$ mode. 

Consider now the Raman signal caused by a single molecule located inside the gap, when the nanocavity position in the excitation field is $\boldsymbol{r}$.  %,\boldsymbol{r'}
For a given excitation frequency $\omega_0$ and polarization $\epsilon=\text{R, A, Lx, Ly}$ (radial, azimuthal, linear along $x$ and $y$, respectively), the Stokes Raman power per unit of input power $P_0$ is proportional to
\begin{equation}\label{eq:Raman}
\frac{P_S(\boldsymbol{r},\omega_0,\epsilon)}{P_0}\propto I_\text{gap}(\boldsymbol{r},\omega_0,\epsilon)R_{zz}^2F_\text{rad}(\omega_S)
\end{equation}
where $R_{zz}=\boldsymbol{n}_z\cdot\boldsymbol{R}\cdot\boldsymbol{n}_z$ is the longitudinal component of the single-molecule Raman polarizability tensor $\boldsymbol{R}$ for the vibrational mode of interest (inside the gap the in-plane field is negligible in comparison to the out-of-plane field); $F_\text{rad}(\omega_S)$ is the radiative Purcell factor, which describes the out-coupling enhancement at the Stokes frequency $\omega_S=\omega_0-\omega_\nu$ corresponding to the vibrational mode at $\omega_\nu$, and it depends on the full spectral response of the plasmonic cavity. The longitudinal field intensity inside the gap is given by \begin{equation}\label{eq:gapfield}
I_\text{gap}(\boldsymbol{r},\omega_0,\epsilon)=I_\text{bg}^z(\boldsymbol{r},\epsilon)K^z(\omega_0)+I_\text{bg}^{xy}(\boldsymbol{r},\epsilon)K^{xy}(\omega_0) 
\end{equation}
where $I_\text{bg}(\boldsymbol{r},\epsilon)=I_\text{bg}^{z}(\boldsymbol{r},\epsilon)+I_\text{bg}^{xy}(\boldsymbol{r},\epsilon)$ is the intensity profile of the focused excitation field $\boldsymbol{E}_\text{bg}(\boldsymbol{r},\epsilon)=\sum_iE^i_\text{bg}(\boldsymbol{r},\epsilon)\boldsymbol{n}_i$ with $i=x,y,z$, such that $I_\text{bg}^{z}(\boldsymbol{r},\epsilon)=|E^z_\text{bg}(\boldsymbol{r},\epsilon)|^2$ and $I_\text{bg}^{xy}(\boldsymbol{r},\epsilon)=|E^{xy}_\text{bg}(\boldsymbol{r},\epsilon)|^2=|E^x_\text{bg}(\boldsymbol{r},\epsilon)|^2+|E^y_\text{bg}(\boldsymbol{r},\epsilon)|^2$ (see Figure~\ref{fig:maps-simul}a,c). 
The longitudinal ($K^z$) and transverse ($K^{xy}$) in-coupling enhancement factors depend on the exact position of the molecule inside the gap (and so does the Purcell factor) and can be estimated from electromagnetic simulations under linearly polarized plane wave excitation. 
When a molecular monolayer occupies the entire gap area, a proper computation of the Raman power should consider collective ``bright" modes \cite{roelli2016} but this more advanced treatment is not necessary here since we do not consider power-dependent collective effects \cite{zhang_optomechanical_2020}. Apart from a different scaling factor, eq.~(\ref{eq:Raman}) remains valid. 

The main conclusion is that eq.~(\ref{eq:gapfield}) describes how the total emitted Raman power depends on the sample position for a given frequency and polarization of the excitation field through the distinct in-coupling enhancement factors for the longitudinal and transverse background field components. 
In particular, if the excitation frequency $\omega_0$ is close to resonance with a longitudinal plasmonic mode, then $K^{z}(\omega_0)\gg K^{xy}(\omega_0)$ and the Raman signal will be strongest when the sample position $\boldsymbol{r}$ matches a maximum of $I_\text{bg}^{z}(\boldsymbol{r},\epsilon)$; conversely, if $\omega_0$ is close to resonance with a transversal plasmonic mode, then $K^{z}(\omega_0)\ll K^{xy}(\omega_0)$ and the Raman signal will be strongest when the sample position $\boldsymbol{r}$ matches a maximum of $I_\text{bg}^{xy}(\boldsymbol{r},\epsilon)$. 
This discussion allows to interpret the experimental confocal Raman maps by connecting them to the near field distribution of the focused excitation beam and the transverse vs. longitudinal nature of the antenna mode.

\paragraph*{Experimental Raman maps of individual nanocavities. ---} 
We excite a nanocube-on-mirror (NCoM) cavity (embedding BPT molecules) with the different light modes introduced earlier: $\boldsymbol{E}_\text{R}, \boldsymbol{E}_\text{A}, \boldsymbol{E}_\text{Lx}, \boldsymbol{E}_\text{Ly}$. Figure~\ref{fig:maps-simul}a shows the calculated local field components produced by these light modes close to the focus of the objective in the $xz$ and $xy$ planes. %, namely $E_\text{loc}^x(\boldsymbol{r},\epsilon)+E_\text{loc}^z(\boldsymbol{r},\epsilon)$ and $E_\text{loc}^x(\boldsymbol{r},\epsilon)+E_\text{loc}^y(\boldsymbol{r},\epsilon)$. 
A Raman intensity map of the $1075$ cm$^{-1}$ vibrational mode excited at $710$~nm (vacuum wavelength) is reported in Figure~\ref{fig:maps-simul}c. By comparing it with the numerically calculated intensity patterns $I_\text{bg}^{xy}(\boldsymbol{r}_0,\epsilon)$ and $I_\text{bg}^{z}(\boldsymbol{r}_0,\epsilon)$ (at the focal plane $z=0$) plotted in Figure~\ref{fig:maps-simul}b we find that for this excitation wavelength Raman scattering improves significantly when the NCoM cavity is positioned in a region where the background excitation field has the largest out-of-plane (longitudinal) component. %It has to be noted that the NCoM has to be positioned according to the specific local pattern of the focused light mode to evidence the improved Raman signals. 
%when the NCOM is positioned in a region where the local focal field is polarized along the $z$ axis.
Indeed, Raman intensity maps collected under radially ($\boldsymbol{E}_\text{R}$) and linearly ($\boldsymbol{E}_\text{Lx}$) polarized illuminations match well the calculated distribution of $I_\text{bg}^{z}(\boldsymbol{r}_0,\epsilon)$ ($\epsilon=\text{R, Lx}$), both in their shapes and relative intensities.
At the chosen excitation wavelength ($\sim710$~nm), our Raman maps thus clearly demonstrate the preferential coupling of the laser field to the longitudinally polarized $J_-$ mode. 

Since the focusing of the azimuthally ($\boldsymbol{E}_\text{A}$) polarized beam results in $I_\text{bg}^{z}(\boldsymbol{r}_0,\text{A})=0$ over the whole focal plane, Raman scattering can then only happen through the excitation of the far detuned $S_{11}$ transverse mode, resulting in a much lower coupling efficiency and weaker Stokes signal. More quantitatively, we found that the ratio between the maximum Raman intensities with radial and azimuthal polarization ($=23.5$) is comparable to the ratio of near-field intensities integrated over the BPT volume with $p$ and $s$ polarization ($=18.5$, Figure~\ref{fig:setup}e).
The polarization of the out-going field is not analysed in these measurements, and it is in general different from the background polarization exciting the nanocavity. We therefore checked that our detection efficiency is largely polarization insensitive for $\sim710$~nm excitation (Figure~\ref{fig:AziHV}), so that we can safely attribute the change of Raman intensity to the change in input coupling efficiency.

Note that the experimental Raman map does not present a full ring pattern as expected for the transverse field of a focused azimuthally polarized beam. A first reason is the unbalanced superposition of $\text{HG}_{10}(x,y)\boldsymbol{n}_y$ and $\text{HG}_{01}(x,y)\boldsymbol{n}_x$ at the back aperture of the objective due to polarization-dependent transmission efficiency of some components (such as the beam splitter). We do measure that the $y$ polarization is attenuated by almost $30\%$ compared to the $x$ component in the excitation path, altering the predicted ring pattern (see Figure~\ref{fig:AziHV}). A second reason is that the nanocavity breaks the cylindrical symmetry and in general exhibits non-degenerate transverse resonances.

%Simulation parameters to reproduce the NCoM nanostructure and the displayed near-field distributions can be found in the SI. 
%This result further confirms that the contribution in excitation to the Raman enhancement, proportional to the square of the local electric field, determines the coupling with the nanostructure.   

Altogether, these first results confirm that Raman intensity maps reveal the polarization-dependent near-field excitation efficiency of localized surface plasmon resonances and can be used to determine the nature of the antenna modes to which the laser preferentially couples at a particular wavelength. 

%%%%%%%%%%%%%%%%%%%%%%%%%%%%%%%%%%%%%%%%%%%%%%%%%%%%%%%%%%%%%%%%%%%%%%%%%%%%%%%%%%%%%%%%%%%%%%%%%%%%%%%%%%%%%%%%%%%%%%%%%%%%%%
\begin{figure}[t!]
\centering
\includegraphics[width=1\textwidth]{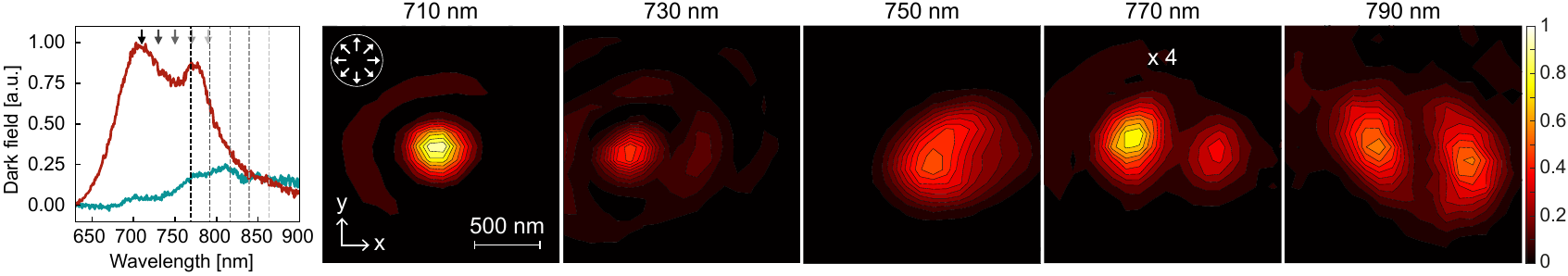}
  \caption{Wavelength sweep under radially polarized illumination. Left panel: Dark-field spectrum of the NCoM when excited with $s$ (turquoise curve) and $p$ (red curve) polarized light. Other panels: Measured maps of the Raman intensity at $1075\SI{}{\centi\meter}^{-1}$ obtained with different excitation wavelengths. Each excitation wavelength is indicated by an arrow in the dark-field plot, while the dashed line with the corresponding grey scale indicates the wavelength of the Raman mode at $\SI{1075}{\centi\meter}^{-1}$.}
  \label{fig:wl-sweep}
\end{figure}
%%%%%%%%%%%%%%%%%%%%%%%%%%%%%%%%%%%%%%%%%%%%%%%%%%%%%%%%%%%%%%%%%%%%%%%%%%%%%%%%%%%%%%%%%%%%%%%%%%%%%%%%%%%%%%%%%%%%%%%%%%%%%%

\paragraph*{Mode-specific coupling.---} 
We now demonstrate how the respective contributions of distinct nanocavity modes to the SERS signals can be identified and modified by tuning the wavelength and polarization of the incoming beam. 
We sweep the excitation wavelength under fixed $\boldsymbol{E}_{R}$ illumination on the same NCoM cavity. As shown in Figure~\ref{fig:wl-sweep}, the single central lobe, consistent with input coupling to the longitudinal $J_-$ mode at $710$~nm, progressively transforms into two lobes as the excitation wavelength increases to $790$~nm. This intensity pattern should be compared with the calculated spatial distribution of $I_\text{bg}^{z}(\boldsymbol{r}_0,R)$ in Figure~\ref{fig:maps-simul}.
Here again, while we expect a ring pattern when the laser couples to the transverse $S_{11}$ mode, the cylindrical symmetry is broken by the nanoparticle shape as well as the polarization dependent transmission on the excitation path and results in a more efficient coupling to a particular transverse polarization, here close to the $x-$axis.  
Complementary data for different incoming light modes and other NCoMs are presented in the SI. 
%It suggests that Raman intensity maps can not only address the coupling factors for different gap modes but also be sensitive to geometrical features of the nano-antenna. 
%From the poor readability of the dark field spectrum under $s$ (purely transverse) polarization shown in Figure~\ref{fig:wl-sweep}, it would have been difficult to make a clear statement on the nature of the cavity modes, while we are able to identify them according to the relative SERS efficiency vs. wavelength.

%%%%%%%%%%%%%%%%%%%%%%%%%%%%%%%%%%%%%%%%%%%%%%%%%%%%%%%%%%%%%%%%%%%%%%%%%%%%%%%%%%%%%%%%%%%%%%%%%%%%%%%%%%%%%%%%%%%%%%%%%%%%%%
\begin{figure}[t!]
\centering
\includegraphics[width=1\textwidth]{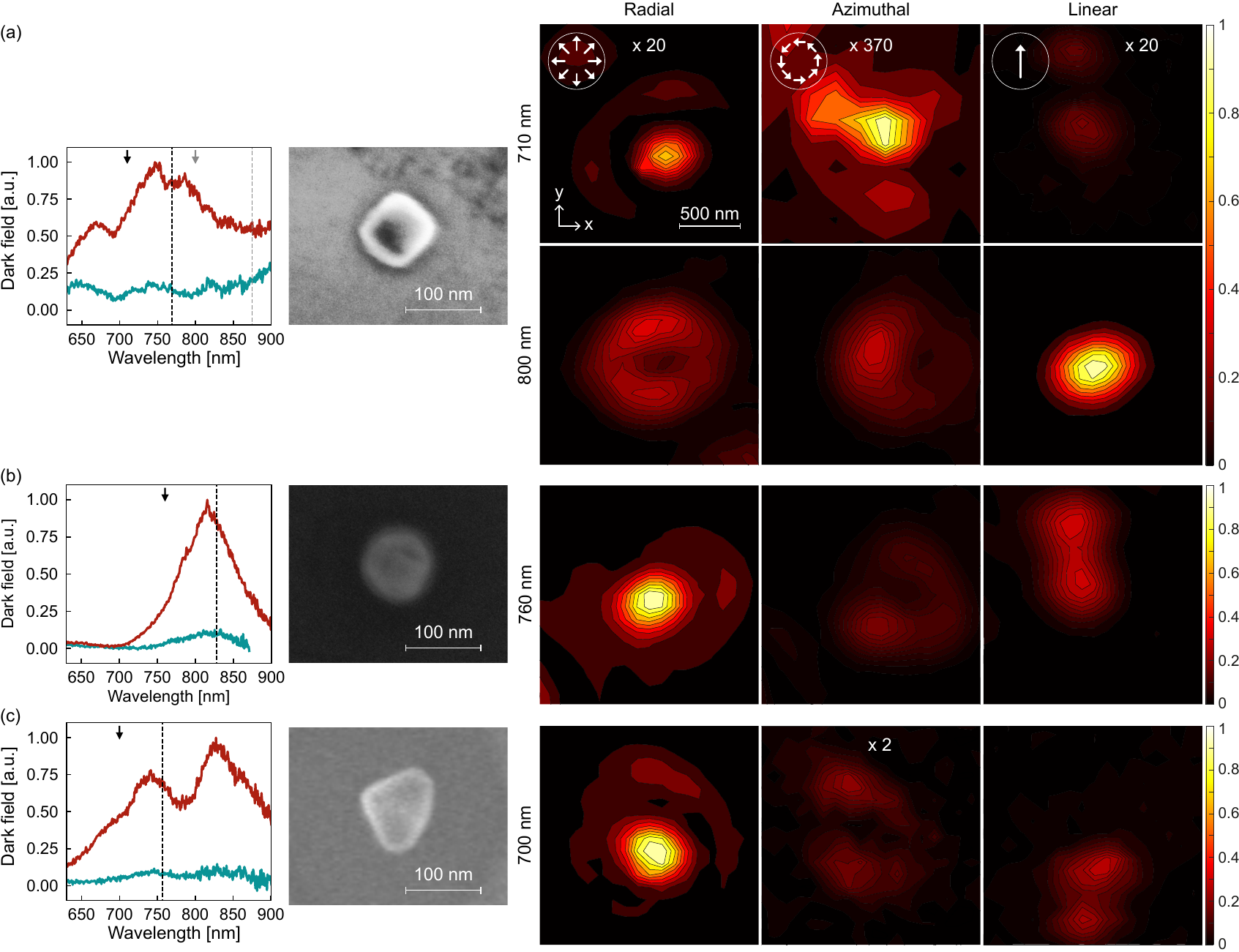}
  \caption{Comparison of different NPoM nanocavities: (a) $75$-nm ALD-coated silver nanocube (NanoXact by nanoComposix), (b) $80$-nm gold nanosphere (by BBI solutions) and (c) triangular gold particle found in a sample of gold nanocubes (A1C-80 by Nanopartz).}
  \label{fig:others}
\end{figure}
%%%%%%%%%%%%%%%%%%%%%%%%%%%%%%%%%%%%%%%%%%%%%%%%%%%%%%%%%%%%%%%%%%%%%%%%%%%%%%%%%%%%%%%%%%%%%%%%%%%%%%%%%%%%%%%%%%%%%%%%%%%%%%

In Figure~\ref{fig:others}, we illustrate the use of our method on nanocavities with various nanoparticle shapes. Panel (a) presents data for another NCoM for which the polarization dependant dark-field scattering spectra is inconclusive as to the nature of the gap modes. 
From the Raman intensity maps, a clear passage from longitudinal to transverse mode coupling is monitored under both radial and linear polarization. In this example, the superiority of the confocal scanning approach compared to polarized dark-field measurements is evident when it comes to determine the nature of the excited modes vs. incoming wavelength.

%Our technique succeeds in the excited mode discrimination where the dark field measurement fails to make clear prediction. Furthermore, even when the dark field is enough to determine the spectral position of the modes, it can't easily establish their nature. For example, 
Theoretical models and numerical studies have predicted how $L_{01}$ and $S_{11}$ modes of quasi-spherical NPoM undergo frequency crossing as a function of the size and shape of the bottom facet of the nanoparticle \cite{kongsuwan_plasmonic_2020, elliott_fingerprinting_2022} -- a parameter that cannot be easily controlled during fabrication nor be measured by existing nanoscopy techniques. Interestingly, our technique can efficiently discriminate these modes' contributions. In the case of the NPoM in panel (b), a single resonance appears in dark field: since the Raman intensity maps at $750$~nm excitation indicates a preferential longitudinal input coupling, we can associate this resonance to the longitudinal $L_{01}$ mode rather than $S_{11}$. Finally, panel (c) corroborates the validity of the technique for a more irregularly shaped NPoM, where we can evidence more efficient coupling to a longitudinally polarized mode around 700~nm. 
It has to be highlighted that our technique could straightforwardly be applied for the characterization of a large variety of nanostructures, including other mirror-based nanocavities (dimer-on-mirror \cite{li2017}, magnetic resonances \cite{chen_plasmon-induced_2018}, etc.) and could be performed using a variety of signals. We illustrate this flexibility in the SI by performing complementary photoluminescence maps of our gold nanocavities under radial and azimuthal illumination.

\section{Conclusion} 
We demonstrated mode-selective laser excitation of nanoparticle-on-mirror systems using a simple technique based on the Raman signal from embedded molecules. 
%In contrast to metal-induced luminescence \cite{}, the Raman signal is sensitive to the field within the NPoM gap, which is of relevance for most quantum and nano-optics applications. 
By performing confocal Raman mapping under radial, azimuthal and linear polarization for different laser wavelengths, we could unambiguously determine the longitudinal vs. transverse nature of the antenna mode most efficiently coupled to the incoming beam and show the transition from one mode to another on a singular nanostructure as the laser wavelength is tuned.
Our method vividly reveals the distinct nature of ultraconfined modes most efficiently excited at particular wavelengths and proves to be a versatile nanoscopy tool able to address irregular nanocavities. Using faster acquisition rates, the method could also be extended to interrogate unstable picocavities \cite{benz2016a}. 
%Other techniques like near-field scanning optical microscopy (NSOM) \cite{yang_mapping_2019, yanai_near-_2014}, fluorescence scanning \cite{novotny_longitudinal_2001}, hyperspectral dark-field (DF) scanning techniques \cite{chikkaraddy2017}, interferometry and polarimetry \cite{rohrich_quantifying_2018} are based on the detection of the light elastically scattered from the nanostructures or from molecules embedded in the gap, and therefore they only probe the far-field radiation, which gives a partial signature of the plasmonic mode. Moreover, we note that fine spatial mapping under linear polarization alone is sufficient to discriminate between dominantly longitudinal vs. transverse mode coupling, making the method particularly easy to implement in most confocal microscopes.
Thanks to its uncomplicated implementation our tool can be combined with other nanoscopy techniques to simultaneously acquire $k$-space \cite{lieb_single-molecule_2004,rohrich_quantifying_2018} or phase information \cite{bauer_nanointerferometric_2014}. Other brighter signals emanating from resonant Raman scattering \cite{haynes_plasmon-sampled_2003,itoh_between_2021} or fluorescence \cite{chikkaraddy2016} could be used as a local probe to learn more and control better their interaction with gap modes, extending on the seminal studies that introduced the technique for focused laser beams \cite{sick_orientational_2000,novotny_longitudinal_2001}.  

This work is a first step toward a precise and quantitative estimate of input coupling rates between far-field radiation and nanocavity modes. These are key quantities in several plasmonic sensing schemes \cite{adato_engineered_2013,neuman_importance_2015} and are essential for accurate modelling and testing of novel predictions of molecular cavity optomechanics, such as the appearance of dynamical backaction and optical-spring effects at the molecular level \cite{roelli2016,jakob_softening_2022}. 
Our results also pave the way for extended studies and optimization of input coupling rates for rapidly evolving hot-spots \cite{benz2016a,lindquist2019,chen_intrinsic_2021} with the help of adaptive-optics illumination \cite{lee_adaptive_2021}. 
More directly, we showed that the control of the incoming beam allows to select the gap mode contributing to the targeted light-matter interaction and enables pushing further the sensitivity of nano-devices using ultra-confined cavity modes.

\section*{Acknowledgements}
This project has received funding from the Swiss National Science Foundation (grants No. 170684, 198898 and 206926) and from the European Union’s Horizon 2020 research and innovation program under Grant Agreement No. 820196 (ERC CoG QTONE).
The authors thank Huatian Hu for assistance with numerical simulations and 
Prof. Kippenberg for offering laboratory access. 

%

%%%%%%%%%% Merge with supplemental materials %%%%%%%%%%
\widetext
\clearpage
\begin{center}
\textbf{\large -- Supplementary Material -- \\Mode-specific Coupling of Nanoparticle-on-Mirror Cavities with Cylindrical Vector Beams}
\end{center}
%%%%%%%%%% Merge with supplemental materials %%%%%%%%%%
%%%%%%%%%% Prefix a "S" to all equations, figures, tables and reset the counter %%%%%%%%%%
\setcounter{equation}{0}
\setcounter{figure}{0}
\setcounter{table}{0}
\makeatletter
\renewcommand{\theequation}{S\arabic{equation}}
\renewcommand{\thefigure}{S\arabic{figure}}
\renewcommand{\thetable}{S\Roman{table}}
\renewcommand{\bibnumfmt}[1]{[S#1]}
\renewcommand{\citenumfont}[1]{S#1}
%%%%%%%%%% Prefix a "S" to all equations, figures, tables and reset the counter %%%%%%%%%%

\paragraph*{FEM simulations. ---} 
In order to understand the results obtained by mapping the Raman signal and confirm the nature of the plasmonic modes that we addressed, we performed FEM simulations with COMSOL Multiphysics (version 6). The simulation geometry is constituted by a $1$-$\SI{}{\micro\meter}$ cubic box, including a gold substrate and a silver nanocube with rounded edges, separated by a molecular gap of height $h_g$. The dielectric functions of gold and silver are interpolations of Johnson's and Christy's data \cite{johnson1972}, while the molecules are modelled as a dielectric layer of refractive index $n_g=1.4$. The rounding factor of the cube's edges is defined by $r=2R/L$, where $L=75$~nm is the total length of the cube and $R$ is the radius of curvature. We used $p$- and $s$-polarized plane-wave excitation with incidence angle $\theta\simeq\frac{\pi}{2}$ from the mirror surface.  

\begin{figure}[b!]
\centering
\includegraphics[width=0.5\textwidth]{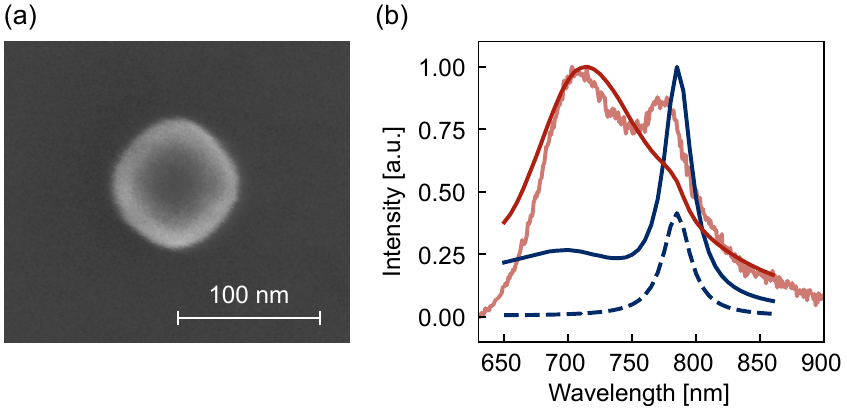}
  \caption{(a) SEM image of the the silver nanocube used for the measurements in Figure~\ref{fig:maps-simul} and \ref{fig:wl-sweep} of the main text. (b) Comparison of the experimental DF intensity with the FEM simulations. In transparency, the DF spectrum of the nanocube when excited with $p$ polarized light is shown (red curve, same as in Figure~\ref{fig:wl-sweep} of the main text). The corresponding FEM simulated scattering cross section is displayed in opaque red. The blue curves represent the simulations of $|E|^2$ integrated over the BPT volume for $p$-polarized (solid line) and $s$-polarized (dashed line) plane-wave excitation (incidence angle $\theta\simeq\frac{\pi}{2}$). The simulations are realized for gap size $2.5$~nm and rounding factor $0.44$.}
  \label{fig:SEM-simul}
\end{figure}

The SEM image in Figure~\ref{fig:SEM-simul}a shows the nanocube used for the measurements in Figure~\ref{fig:maps-simul} and \ref{fig:wl-sweep} 
of the main text. We first performed DF measurements in order to locate the main plasmonic resonances. Then, we implemented FEM simulations of the scattering cross section with parameters $h_g=2.5$~nm and $r=0.44$ chosen to match the measured frequencies of the DF peaks, as shown in Figure~\ref{fig:SEM-simul}b. 

The gap height is determined by the length of the molecules and their orientation, and was shown to increase for longer incubation time of the BPT SAM \cite{ahmed2021}. The BPT molecule have a length of $1.4$~nm, while the PVP ligand have lengths around $5$~nm but tend to lie flat around the nanocube; therefore, $h_g=2.5$~nm appears to be a reasonable choice. A precise estimation of the rounding parameter is not less difficult, since the major contribution to the gap mode distribution comes from the bottom facet of the cube, which is hidden to SEM imaging. Furthermore, the SEM picture shows the top facet after ALD coating: on the one hand the coating slightly reshapes the silver cube, so that a picture before coating would not be illustrative of the measured nanostructure; on the other hand, ALD fails to strictly reproduce a curved surface because of its spongy nature \cite{chen2018a}. As a consequence, while the rounding factor estimated from the SEM images of the producer (nanoComposix) is around $0.2$, Figure~\ref{fig:SEM-simul}a shows rounding factors between $0.7$ and $0.8$ for a total length of $85$~nm. The latter is in accordance with the presence of a $75$-nm silver nanocube with $5$-nm ALD coating. Considering a $10\%$ less in the value of $r$ for the actual silver structure and given the above mentioned complications, the choice of $r=0.44$ in the simulation is realistic.
%If the ALD was perfectly reproducing the corner's shape, the $r$ parameter of the silver structure could be calculated to be around $10\%$ less than the measured one. We subtract another $40\%$
%$r=r'-\frac{a}{L+2a}$ with $a=5nm$

In Figure~\ref{fig:SEM-simul}b, we also show the integrated near-field intensity for the two linear polarizations. The integration volume is $V=\SI{1}{\micro\meter}\times\SI{1}{\micro\meter}\times h_g$, occupied by the molecular spacer. As expected \cite{zuloaga_energy_2011}, the mode resonances are slightly shifted ($\lesssim 15$~nm) with respect to the scattering curves.   

\paragraph*{Setup details. ---}
The excitation beam is provided by an MBR-01 Ti:Sapph ring cavity laser by Coherent, optically pumped by a Verdi-V10 532~nm diode-pumped solide-state laser (also by Coherent). The wavelength of the Ti:Sapph can be tuned from $700$~nm to $1000$~nm. 
We set the polarization with a nematic liquid-crystal device, namely a radial polarization converter by ARCoptix. It is equipped with an electrical driving (LC Driver by ARCoptix) in order to convert the incoming linearly polarized beam into a radially or azimuthally polarized beam.

In order to map the Raman intensity along the sample surface, we place the sample on a piezo nanopositioner Nano-Align5-100 from Mad City Labs, Inc. It is actuated in closed loop and it has a resolution of $0.2$~nm in $x$, $y$ and $z$ directions. Notice that the final resolution of the measured Raman intensity maps is also affected by  mechanical drifts of the sample in the three directions, especially because a single map requires more than $10$ minutes of acquisition time. This explains why the Raman maps with two lobes usually show higher intensity from one of them, in particular the lobe close to the center of the figure: the maps are $xy$ automatized scans taking the center of the figure as starting point. Before the acquisition, the starting point is optimized to obtain maximum detected Raman signal, while, during the acquisition, the sample can slightly shift out of the focal plane.

\begin{figure}[]
\centering
\includegraphics[width=0.5\textwidth]{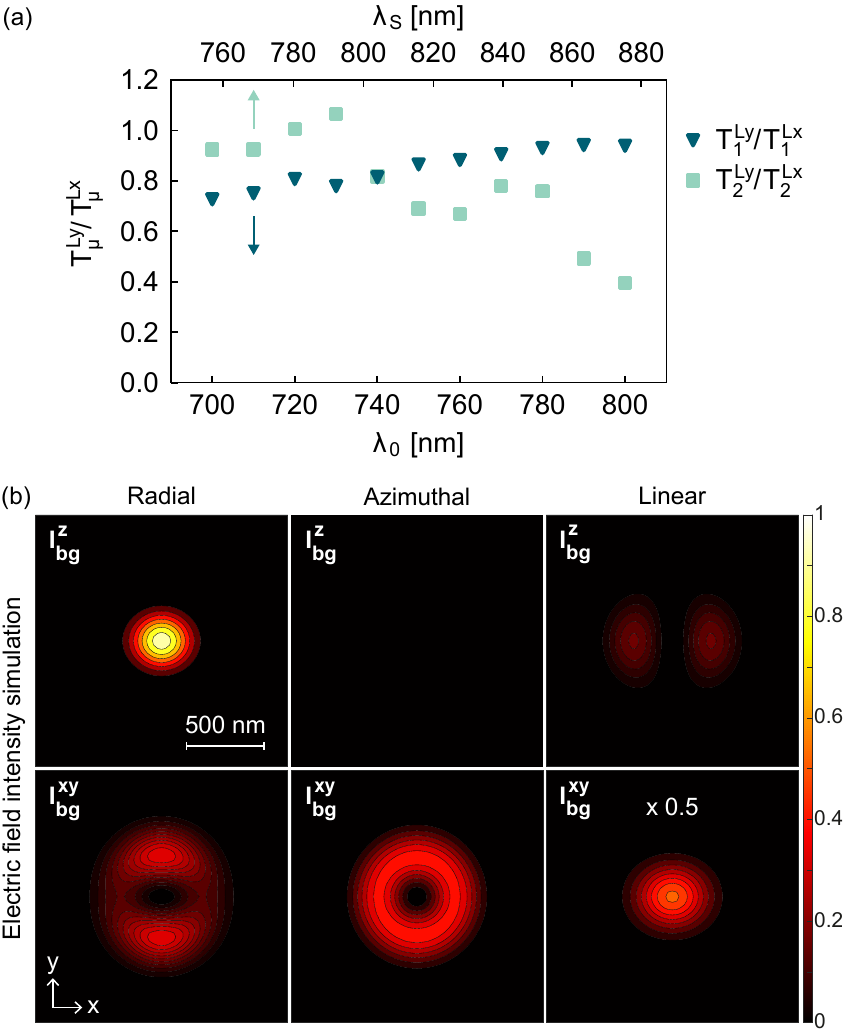}
  \caption{Effect of unequal transmission of the $x$ and $y$ polarization components. (a) Plot of the $y$-$x$ ratio of the transmission factors as a function of wavelength: $T^\text{Ly}_1/T^\text{Lx}_1$ is represented by triangles, $T^\text{Ly}_2/T^\text{Lx}_2$ by squares ($\mu=1,2$). (b) Longitudinal ($z$) and transverse ($r$) components of the focused field intensity $I_\text{bg}(\boldsymbol{r}_0,\epsilon)$ (across the focal plane $z=0$), calculated for an excitation wavelength of $710$~nm and a factor $0.7$ of polarization unbalance. Notice that, with such an unbalance, the azimuthal beam acquires a tiny longitudinal component (a factor 160 less intense than the radial $I^z_{bg}$), which we don't show here.}
  \label{fig:AziHV}
\end{figure}
%Raman maps obtained with azimuthally polarized excitation at $760$~nm and $780$~nm and a polarizer in the detection path. The arrow on the right corner of each panel indicates the orientation of the polarizer. The data are normalized for $P_1^iT_2^i$. SEM picture of the nanoparticle used can be found in Figure~\ref{fig:full_scan}

\textit{Setup calibration. ---}
All the presented Raman measurements are normalized for the setup response. The calibration involves two steps: (i) for each excitation wavelength $\lambda_0$, the power at the sample position $P_2(\lambda_0)$ has been measured as a function of the reference power $P_1(\lambda_0)$, which is recorded during the measurements;  (ii) for each Stokes wavelength $\lambda_S\equiv\lambda_S(\lambda_0)$, the Ti:Sa has been tuned to $\lambda_S$ and the intensity $I(\lambda_S)$ has been acquired with the spectrometer as a function of the power $P_2(\lambda_S)$. Each measurement has been performed with linearly polarized illumination along $x$ and $y$. If $S^{\epsilon}(\lambda_S)$ is the acquired Stokes intensity, where $\epsilon=\text{R, A, Lx, Ly}$ indicates the excitation polarization, the calibrated intensity is given by
\begin{equation}
    S_\text{cal}^{\epsilon}(\lambda_S) = \frac{S^{\epsilon}(\lambda_S)}{P_1^{\epsilon}(\lambda_0)}\frac{1}{T^\epsilon_1(\lambda_0)}\frac{1}{T^\text{ave}_2(\lambda_S)}
\end{equation}
where we defined the transmission factors $T^\epsilon_1(\lambda_0)=P_2^{\epsilon}(\lambda_0)/P_1^{\epsilon}(\lambda_0)$ and $T^\text{ave}_2(\lambda_S)=(T^\text{Lx}_2(\lambda_S)+T^\text{Ly}_2(\lambda_S))/2$, with $T^\epsilon_2(\lambda_S)=I^{\epsilon}(\lambda_S)/P_2^{\epsilon}(\lambda_S)$. For radial and azimuthal illumination, we assumed $T^\text{R}_1(\lambda_0)=T^\text{A}_1(\lambda_0)=(T^\text{Lx}_1(\lambda_0)+T^\text{Ly}_1(\lambda_0))/2$. Note that $S_\text{cal}^{\epsilon}(\lambda_S)$ corresponds to the efficiency of the Raman process.

\begin{figure*}[]
\centering
\includegraphics[width=1\textwidth]{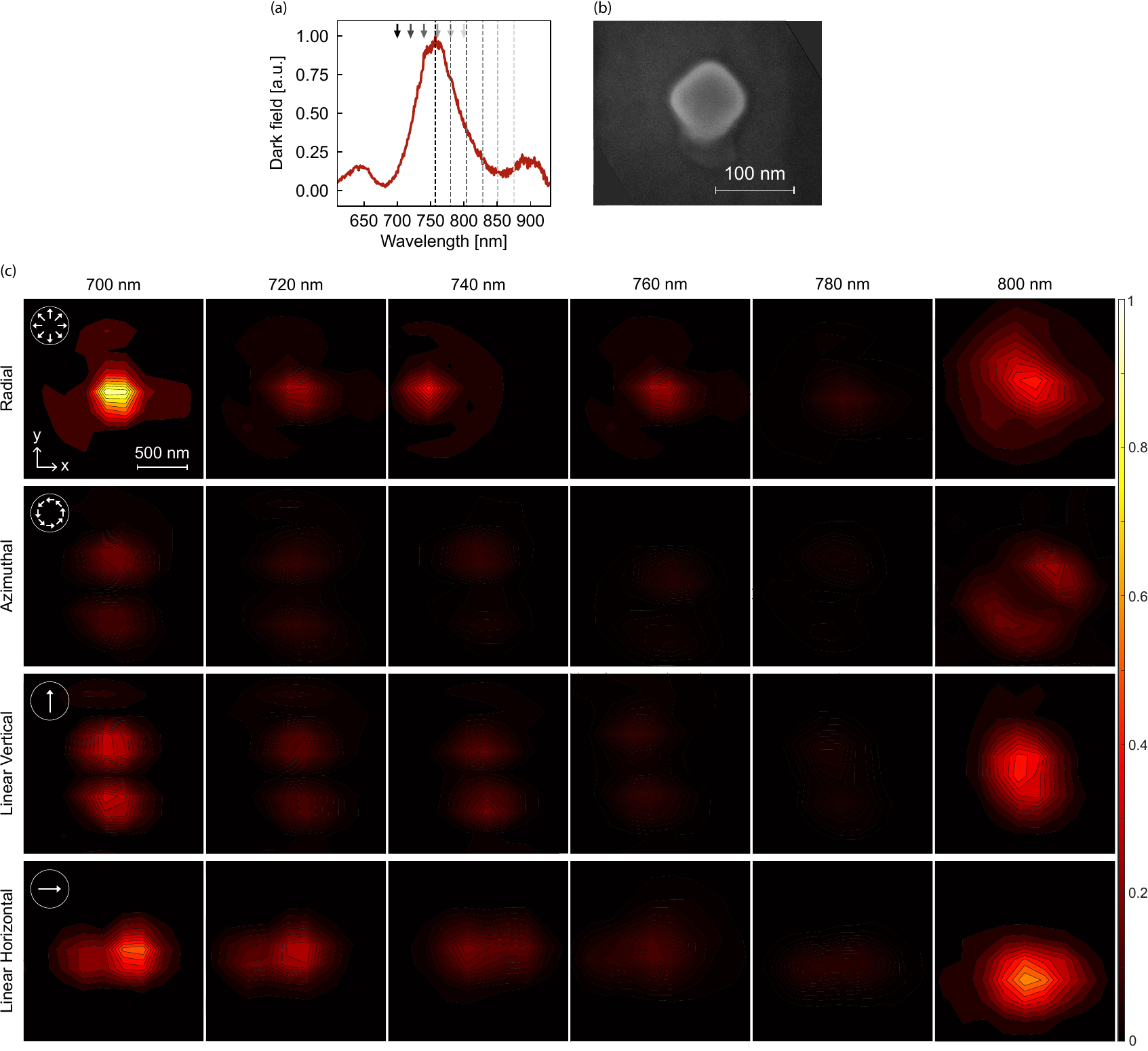}
  \caption{Complete wavelength and polarization sweep on a silver nanocube. (a) DF spectrum of the nanocavity when excited with $p$ polarized light. (b) SEM image of the nanocube. (c) Raman intensity maps for different wavelengths and polarizations. Each wavelength is indicated as an arrow in panel (a), while the dashed line with the corresponding colour indicates the wavelength of the Raman mode at $\SI{1075}{\centi\meter}^{-1}$.}
  \label{fig:full_scan}
\end{figure*}

We notice that, when the transverse mode is excited, most of our Raman maps present a stronger intensity of the $x$ polarization component, so that two intensity lobes appear instead of a doughnut. In fact, when the excitation beam is azimuthally polarized, the two lobes appear along the $y$ axis (see Figure~\ref{fig:maps-simul}b and Figure~\ref{fig:full_scan}), while they appear along $x$ for radial polarization (see Figure~\ref{fig:wl-sweep}). This is due to the fact that the two ortoghonal polarizations are not equally transmitted through the setup, as shown in Figure~\ref{fig:AziHV}a, where the $y$-$x$ ratios of the transmission factors are plotted as functions of the excitation and Stokes wavelengths. The $x$ polarization results more efficiently transmitted: the polarization unbalance in excitation is lower than $30\%$, while in the detection it reaches up to $60\%$. In Figure~\ref{fig:AziHV}b, the longitudinal and transverse components of the focused field intensity are calculated by including a polarization unbalance factor of 0.7.  
%The effect of $T_2^\epsilon$ on the measured intensity is well displayed by Figure~\ref{fig:AziHV}(b), where the Raman signal is acquired with azimuthally-polarized excitation and a polarizer in the detection path. Even though the distribution of the Raman intensity is determined in excitation and it doesn't change by rotating the polarizer, the measured intensity with a $y$ polarizer is attenuated by more than a factor 4. % 4.2 for 760, 4.5 for 780. i.e. 76% at 760nm, 78% at 780nm

In few cases the preferential axis is not clearly defined, as in Figure~\ref{fig:others}(a-b). We think that in these cases a polarization unbalance in the excitation beam played a major role while summing up to the effect of the unequal $x$-$y$ transmission. In fact, the quality of the doughnut beam, generated by the coupled effect of the LC device and the spatial filter, is very sensitive to alignment, and difficult to control while changing the Ti:Sa wavelength. Typical beam profiles, as the one shown in Figure~\ref{fig:setup}b, present an axis of higher intensity.

%\begin{comment}
\begin{figure}[h!]
\centering
\includegraphics[width=0.7\textwidth]{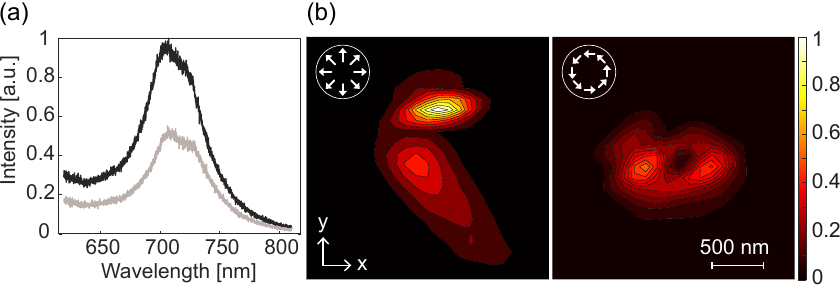}
  \caption{PL measurements. (a) Spectra acquired at the position of maximum intensity under $532$-nm excitation with radial (black line) and azimuthal (gray line) polarization. The acquisition time is $5$~s. (b) PL intensity maps for radially and azimuthally polarized excitation. For each $xy$ position, we plot the integrated intensity over the entire spectral range.}
  \label{fig:PL}
\end{figure}
\textit{Photoluminescence maps.---}
In order to perform photoluminescence (PL) measurements, the experimental setup has been complemented with an additional path enabling excitation at $532$~nm with radial and azimuthal polarization. The PL spectra and the $xy$ maps obtained for the two polarization are shown in Figure~\ref{fig:PL}. Both spatial distributions evidence the coupling of the in-plane components of the field with similarintensities and spectral profiles, confirming that the 532-nm excited PL is assisted by the transversal plasmonic resonance of the nanoparticle \cite{lumdee_gap-plasmon_2014}.
%\end{comment}

\textit{Additional data.---} 
For completeness of the present work, additional figures are reported in this section. 

\begin{figure*}[]
\centering
\includegraphics[width=1\textwidth]{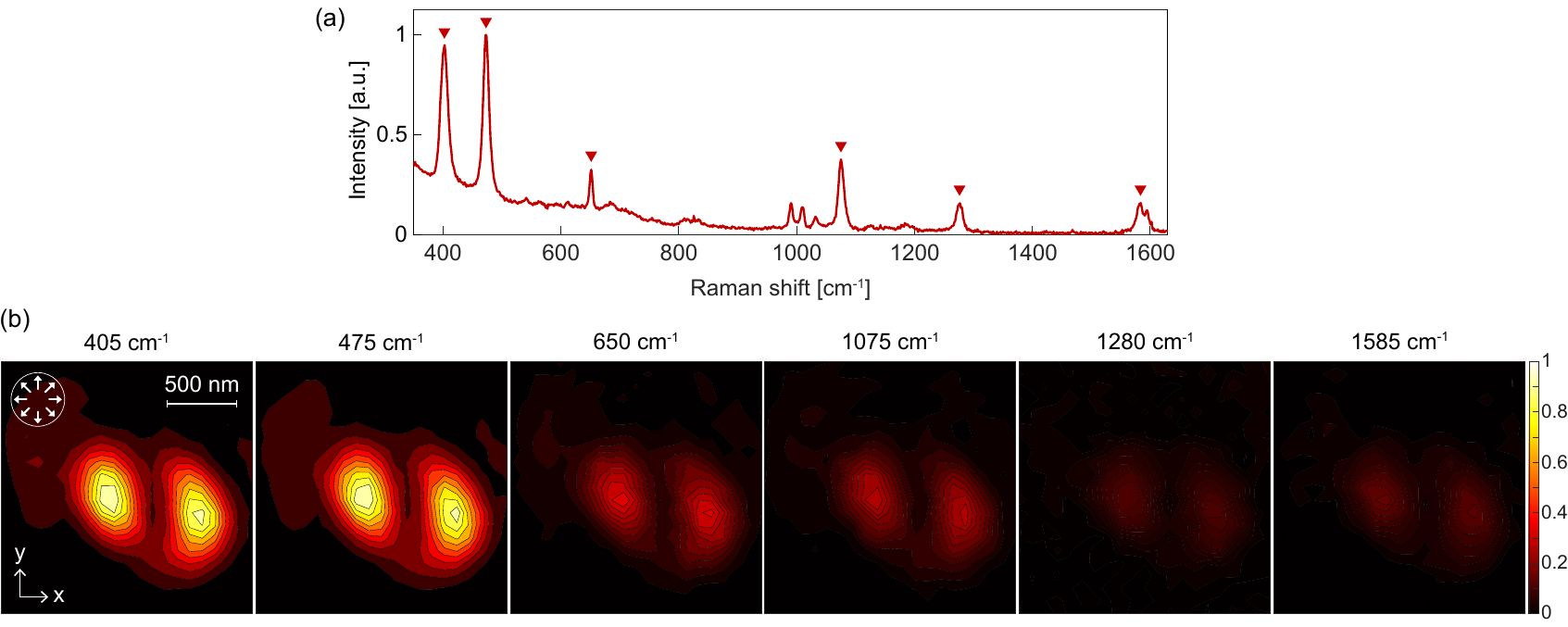}
  \caption{Comparison of Raman maps for different vibrational modes. (a) Stokes spectrum in the position of maximum intensity, obtained by exciting a nanocube with a $790$-nm radially-polarized beam. The most prominent peaks are indicated by red triangles. (b) $xy$ intensity maps of the selected Stokes peaks.}
  \label{fig:shifts}
\end{figure*}
In Figure~\ref{fig:shifts} we show that the intensity maps do not change for different choices of the Stokes peak.
%In the case of the BPT molecule, the Raman tensor of different modes are very similar, so that it's difficult to draw conclusions on the role of the Raman tensor

\begin{figure}[]
\centering
\includegraphics[width=0.6\textwidth]{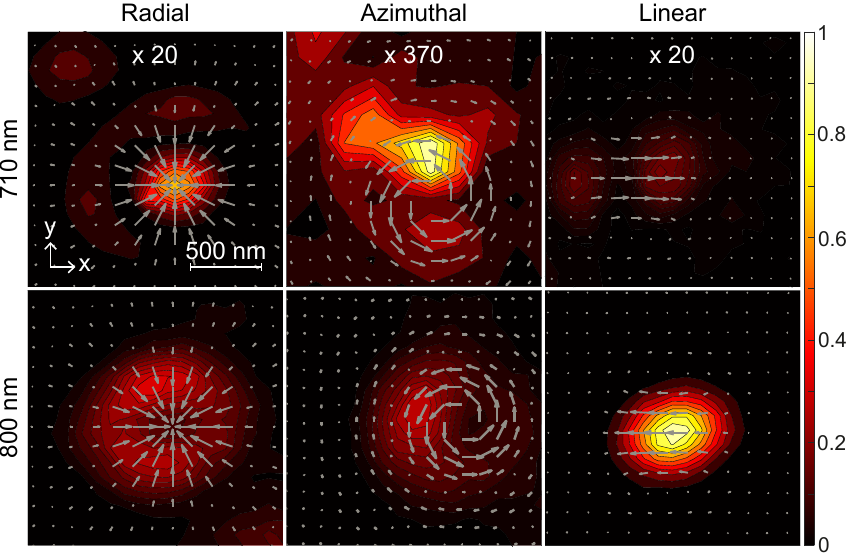}
  \caption{Direct comparison of the Raman maps in Figure~\ref{fig:others} of the main text with the numerically calculated excitation field in the focal plane (grey arrows).}
  \label{fig:Esuperimp}
\end{figure}
In order to have a more direct comparison between the calculated focused field distribution and the measured Raman maps, we superimposed them in Figure~\ref{fig:Esuperimp} while keeping the same scale.

Furthermore, we performed a wavelength sweep along the $J_-$ mode of the nanocube in Figure~\ref{fig:full_scan}a for radial, azimuthal and two orthogonal linear polarizations. 

\begin{figure}[]
\centering
\includegraphics[width=0.68\textwidth]{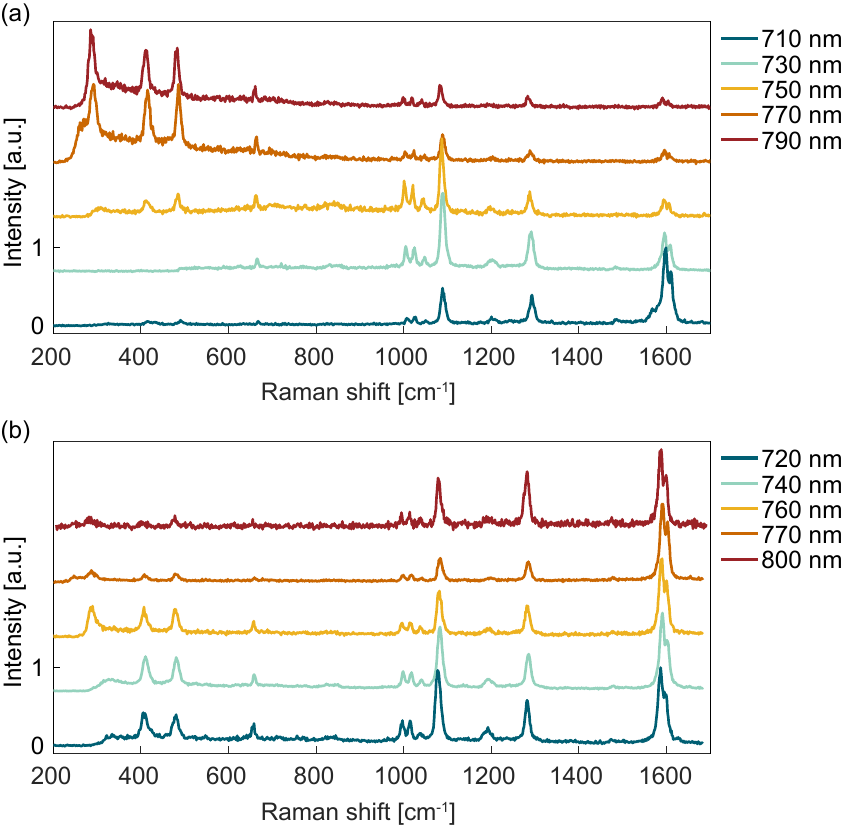}
  \caption{Effect of the plasmonic response on wavelength-dependent SERS measurements for (a) the nanocube used in Figure~\ref{fig:wl-sweep} (b) the nanocube depicted in Figure~\ref{fig:full_scan}a. The wavelength scans were performed under radially polarized illumination. Stokes spectra were selected on Raman maps at the position of their maximum intensity for different excitation wavelengths (indicated in the legend). Each spectrum is calibrated for the setup response and normalized. Tunable interference filters were adjusted to cover adequately the spectral region of interest.}
  \label{fig:cascade}
\end{figure}

Finally, we evidence the effect of the plasmonic response on the spectra acquired from different nanocubes in Figure~\ref{fig:cascade}. The spectral response of the antenna shapes both the in-coupling and the out-coupling efficiencies. As a result, the relative enhancement of Raman peaks varies between spectra measured at different excitation wavelengths.   

\end{document}